\documentclass[10pt]{article}

\usepackage{amsmath}
\usepackage{amsfonts}
\usepackage{amssymb}
\usepackage{amsthm}
\usepackage{stmaryrd} 
\usepackage{braket}
\usepackage{empheq}
\usepackage{graphicx}
\usepackage{subfigure}
\usepackage[round]{natbib}
\usepackage{footmisc}
\usepackage{appendix}
\usepackage{epsf}
\usepackage{epsfig}
\usepackage{tikz,tikz-3dplot}
\usetikzlibrary{decorations.markings,decorations.pathmorphing}
\usepackage{caption}
\usepackage{pgfplots}
\usepackage[utf8]{inputenc}
\usepackage[T1]{fontenc}
\usepackage[english]{babel}
\usepackage{sectsty}
\usepackage{hyperref}

\subsubsectionfont{\normalfont\itshape}
\usepackage{fullpage}

\renewcommand{\bar}{\overline}
\renewcommand{\leq}{\leqslant}

\DeclareMathOperator{\tr}{tr}

\theoremstyle{plain}
\newtheorem{theorem}{Theorem}[section]

\theoremstyle{definition}
\newtheorem{definition}[theorem]{Definition}

\title{The measurement problem in the light \\ of the theory of decoherence}
\author{\textsc{Antoine Soulas} \\ Institut de recherche mathématique de Rennes (IRMAR) \\ Université de Rennes and CNRS, Rennes, France  \\  \small{Email: antoine.soulas@univ-rennes1.fr}}

\begin{document}
\maketitle

\abstract{Endeavoring to formulate an exhaustive solution to the measurement problem in view of the theory of decoherence leads to a better understanding of the status of the collapse and of the emergence of classicality, thanks to a precise definition of the measurement and some new vocabulary to speak about quantum mechanics. Considering the latter as a probabilistic theory all along allows us to avoid the usual probability problem of the many-worlds interpretations. A thorough verification of the consistency of quantum mechanics at all scales is proposed, as well as a discussion of what can be deemed an observer.}

\paragraph{Keywords:}measurement problem, interpretation of quantum mechanics, decoherence, special relativity, locality, determinism.

\section*{Acknowledgements}
I would like to gratefully thank my PhD supervisor Dimitri Petritis for the great freedom he grants me in my research, while being nevertheless always present to guide me. I also thank my friends Dmitry Chernyak and Matthieu Dolbeault for illuminating discussions.

\section*{Statements and Declarations}
\textbf{Funding:} no funding was received for conducting this study. \\
\textbf{Competing Interests:} the author has no competing interests to declare that are relevant to the content of this article.\\
\textbf{Ethics approval, Consent, Data, Materials and code availability, Authors’ contribution statements:} not applicable. 


\newpage
\section*{Introduction} \label{intro}

Measuring a quantum system affects its physics: the quantum Zeno effect or the fact that the presence of a detector in a double slit experiment destroys the fringes confirms this undoubtedly. The measurement must therefore be treated as a physical evolution. Apparent conceptual problems concerning the implementation of measurements in quantum mechanics (QM) fall into three main categories: 

\begin{enumerate}
\item How comes that the wave function collapse is a non-linear and probabilistic evolution, whereas the Schrödinger equation is linear and deterministic?
\item What is a measurement? What is an observer? When exactly does the projection occur? If it exists, where is the border between the classical and the quantum world? If there is none, how does classicality emerges from the quantum? This is the usual \textit{measurement problem of QM}\footnote{It is impressive to realize how uncertain the physical community still remain on this fundamental problem. A poll carried out by \cite{schlosshauer2013snapshot} among the participants of a conference on the foundations of QM in 2011 revealed that, when asked about the measurement problem, the participants were approximately equally divided between the five possible answers: a pseudoproblem, a problem solved by decoherence, a problem solved/will be solved in another way, a severe difficulty threatening QM, none of the above. A later survey \citep{sivasundaram2016surveying}, gathering more participants in different universities, also led to very divergent answers. Besides, in a recent interview \citep{interviewpenrose}, Penrose described the measurement problem almost exactly as it was already formulated a century ago, considered it as an inconsistency of quantum theory (mainly because of the absence of clear definition of what constitutes a measurement), and deplored the fact that most physicists ‘shove [this problem] under the carpet’.}.
\item How to model the effect of a measurement in a way compatible with special relativity? Indeed, the measurement is generally described as an update happening instantaneously in the whole space, but in special relativity the notion of instantaneity is not defined.
\end{enumerate}

The aim of this paper is to formulate properly and propose a solution to the first two problems (the third one is tackled in another article \cite{soulas2023logical}). The reader expecting that the solution to the measurement problem relies on a mathematical feat, or on a stunning simple idea no one ever had, will be disappointed. It rather consists of cautiously changing the way we think and speak of QM. Eventually, the interpretation we develop could perhaps be seen as a many-worlds interpretation with more philosophical rigour, where we are very careful not to give more ontology to the ‘worlds’ than they deserve, so that the problem of the meaning of the probabilities evaporates. To deliver or to receive a new language game, as understood by \cite{wittgenstein1953philosophical}, is not an easy task, so the reader must be warned that this section might appear slightly recondite on a first reading. 

We investigate methodically the measurement problem of QM in all its ramifications, in view of the ideas of the theory of decoherence. We don't want to be content with a list of several attempts of solution and their respective drawbacks. Our ambitious aim is rather to understand the relation between the different pieces of the puzzle and propose an exhaustive solution, as well as some linked considerations that may be worth sharing. Since the different issues are all very entangled, we will refer a lot, throughout the reasoning, to previous or later paragraphs.

After some simple preliminaries on quantum theory and decoherence (\S\ref{preliminaries}), we decompose the problem into two of its main formulations: the status of the collapse (\S\ref{first_formulation}) and the disappearance of (most) quantum effects at our scales (\S\ref{second_formulation}). In the first part, we argue that the collapse is not a physical process (\S\ref{epistemic/ontic}, \S\ref{physical_collapse}), explain its nature and validity, give a precise definition of the measurement and develop some new vocabulary to interpret QM (\S\ref{definition_measurement}). In the second, we introduce some ideas based on Schrödinger's cat (\S\ref{Schrödinger's cat}) further pursued in the sequel, where we tackle the preferred-basis problem and explain the disappearance of interferences (\S\ref{preferred_basis}), of quantum correlations due to entanglement (\S\ref{correlations}) and of indeterminism (\S\ref{determinism}). We finally check that the assumption of the universal validity of QM has no internal inconsistencies (\S\ref{consistency}) and discuss the notion of observer (\S\ref{observer}).

\section{Preliminaries} \label{preliminaries}
\subsection{Preliminary remarks on quantum mechanics} \label{remarks on QM}

Let's begin with some obvious but crucial remarks about the nature of quantum theory that are so easily forgotten, thereby often leading to confusion and headaches. Unlike deterministic classical physics,
\begin{enumerate}
\item \textit{QM is a probabilistic theory}\footnote{Einstein thought that QM was incomplete because of his famous ‘God does not play dice’. However, note how positivist this remark is: it is not that God necessarily plays dice, but only that the human mind may not be able to access better than statistically the way God plays.} that predicts the statistics of empirical facts observed following measurements,
\item in QM, the act of measuring a system affects its future statistics because (i) it creates entanglement and therefore produces decoherence, (ii) it requires to interact with the system which perturbs it (see \S\ref{determinism}).
\end{enumerate}

It is the combination of these two ingredients that makes almost impossible to built an ontological picture for quantum physics. Indeed, the ‘actual state’ of a particle (\textit{e.g.} through which slit did the electron pass?) can not be based on the ground of its physical state, the density matrix, because the latter is not deterministic and therefore doesn't provide a unique, well-defined answer. Moreover, it can not be based either (as can be done, for example, in statistical physics) on ‘the outcome that one would have obtained if one had measured’, because then the system would not be the same anymore and the future evolution of the particle would have been different. 

The best attempt to assign realistic ontological properties to particles is the fascinating De Broglie - Bohm interpretation, but:
\begin{itemize}
\item it is highly non-local (as would any hidden variable theory, due to Bell's theorem) and therefore its ‘realistic’ trajectories can actually lead to very surrealistic phenomena \citep{englert1992surrealistic}. In particular, a ‘detector in an interferometer (...) can become excited even when the electron passes along the other arm of the interferometer’ \citep{dewdney1993late}, while the detector through which the electron actually passes remains unexcited. Consequently, Bohmian trajectories don't represent either ‘the outcome that one would have obtained if one had measured’. What kind of interesting meaning can they possibly have, then? Besides, does the very notion of physics still makes sense when accepting that we can not trust our measurement devices? 
\item no one has yet succeeded to make it compatible with special relativity,
\item because of item 2.\ (ii), it is impossible to know perfectly both the initial position and momentum of a particle, therefore it is still a probabilistic theory that adds no predictive power to standard QM.
\end{itemize}

\subsection{Selective measurement, non-selective measurement} \label{selective/non-selective}
When discussing the implementation of measurements in QM, it is crucial to distinguish clearly between selective and non-selective measurement. The former is what is usually referred to as the ‘wave function collapse’. When measuring an observable $\hat{A}$, of spectral decomposition $\hat{A} = \sum_{x \in \mathrm{spec}{\hat{A}}} x \Pi_{x}$, relative to a system in a pure state (resp.\ mixed state) $\ket{\Psi}$ (resp.\ $\rho$), if the outcome is the eigenvalue $x_0$, then the system's state is standardly postulated to evolve as: 

\begin{equation} 
\ket{\Psi} \longrightarrow \frac{\Pi_{x_0} \ket{\Psi}}{\lVert \Pi_{x_0} \ket{\Psi} \rVert } \quad \quad \text{resp.} \quad \rho \longrightarrow  \frac{\Pi_{x_0} \rho \Pi_{x_0}}{ \tr(\rho \Pi_{x_0}) }. \label{selective}
\end{equation}
Note that the implementation of such a selective measurement requires to know (or assume) the outcome. Alternatively, a non-selective measurement describes the update of the state without distinguishing the actual outcome: it is only concerned about the statistics that would be obtained if the experiment were repeated. Except for specific cases, the result is always a mixed state: 

\begin{equation}
\rho \longrightarrow \displaystyle \sum_{x \in \mathrm{spec}{\hat{A}}} \Pi_{x} \rho \Pi_{x}. \label{non-selective}
\end{equation}
Selective measurement is a non-linear probabilistic operation on $\rho$, whereas non-selective measurement is a linear deterministic one, since it merely consists of extracting the diagonal part of the density matrix in the eigenbasis of the measured observable. The theory of decoherence explains how non-selective measurements \eqref{non-selective} arise in QM, due to the entanglement between the system and its measurement apparatus. Let's briefly recall why, following our presentation given in more detail in \citep{soulas2023decoherence}.

\subsection{Basic notions of decoherence} \label{basics}

According to \cite{di2021stable}, the deep difference between classical and quantum is the way probabilities behave: all classical phenomena satisfy the total probability formula 
\[ \mathbb{P}(B=y) =  \sum_{x \in \mathrm{Im}(A)} \mathbb{P}(A=x) \mathbb{P}(B=y \mid A=x) \]
relying on the fact that, \textit{even though the actual value of the variable $A$ is not known, one can still assume that it has a definite value among the possible ones}. This, however, is not correct for quantum systems, for which the diagonal elements of their density matrix account for the classical behavior (they correspond to the terms of the total probability formula) while the non-diagonal terms are the additional interference terms\footnote{As a reminder, this is because the probability to obtain an outcome $x$ is:\\  $\tr(\rho \ket{x}\bra{x}) = \sum_{i,j=1}^n \rho_{ij} \braket{j \vert x}  \braket{x \vert i} = \sum_{i=1}^n \underbrace{\rho_{ii}  \vert \braket{x \vert i} \vert^2}_{\mathbb{P}(i) \mathbb{P}(x \mid  i)}  +  \sum_{1\leq i < j \leq n}  \underbrace{2 \mathrm{Re}( \rho_{ij}  \braket{j \vert x} \braket{x \vert i })}_{\text{interferences}}$ }. The aim of the theory of decoherence is to understand why the latter go to 0 for non-isolated systems. 

Here is the typical situation. Consider a system $\mathcal{S}$, described by a Hilbert space $\mathcal{H}_{\mathcal{S}}$ of dimension $d$, that interacts with an environment $\mathcal{E}$ and let $\mathcal{B} = (\ket{i})_{1\leq i \leq d}$ be an orthonormal basis of $\mathcal{H}_{\mathcal{S}}$. In the sequel, we will say that each $\ket{i}$ corresponds to a \textbf{possible history} of the system in this basis. In these terms, QM is nothing but a strange geometrical way of calculating probabilities in which all the possible histories interfere, according to the non-diagonal part of the density matrix. For simplicity, let's also assume for now that $\mathcal{B}$ is a \textbf{conserved basis} during the interaction with $\mathcal{E}$ (in some contexts called a pointer basis) and denote $\ket{\Psi} = \left( \sum_{i=1}^d c_i \ket{i} \right) \otimes \ket{\mathcal{E}_0}$ the initial state of $\mathcal{S} + \mathcal{E}$. After a time $t$, the total state evolves to $\ket{\Psi(t)} = \sum_{i=1}^d c_i \ket{i} \otimes \ket{\mathcal{E}_i(t)}$. Define $\eta_{\mathcal{B}}(t) = \displaystyle \max_{i \neq j} \; \lvert \braket{\mathcal{E}_i(t) \vert \mathcal{E}_j(t)} \rvert$. Note that this definition actually makes sense in any (potentially non-conserved) basis. We will drop the subscript $\mathcal{B}$ when the context is clear. By taking the partial trace with respect to the environment, it is straightforward to see that the i$^{th}$ diagonal coefficient of the system's state $\rho_{\mathcal{S}}$ always remains equal to $\lvert c_i \rvert ^2$, whereas the non-diagonal ones are $c_i \bar{c_j} \braket{\mathcal{E}_j(t) \vert \mathcal{E}_i(t)}$, bounded in modulus by $\eta(t)$. Therefore $\eta$ \textit{measures how decohered the system is}, that is how close it is from satisfying the total probability formula. It is now well understood why $\eta$ quickly tends to 0 for some interesting bases in a wide variety of contexts \citep{zurek2003decoherence,joos1996decoherence}. In particular, when $\mathcal{E}$ is a measurement apparatus for the observable $A$, the eigenbasis of $\hat{A}$ is clearly a conserved basis\footnote{This is how a measurement works: if $\mathcal{S}$ is prepared in an eigenstate of $\hat{A}$, it should remain in this state during the measurement of $\hat{A}$. In general, the existence of a conserved basis is not guaranteed unless the interaction Hamiltonian takes the form $\sum_i \Pi^{\mathcal{S}}_i \otimes H^{\mathcal{E}}_{i}$, where $(\Pi^{\mathcal{S}}_i)_{1 \leq i \leq d}$ is a family of orthogonal projectors associated to this basis.}, yielding the non-selective measurement evolution \eqref{non-selective} in the ideal case of perfect and immediate decoherence. 

\section{First formulation: the transition from mixed to pure states} \label{first_formulation}
\subsection{Epistemic vs.\ ontic meaning of the wave function} \label{epistemic/ontic}

We are now ready to tackle the measurement problem. The usual puzzle is to make sense of the fact that only one of the possible histories is actually observed, and to understand when and why the state evolves from $\rho_{\mathcal{S}} =  \lvert c_1 \rvert ^2 \begin{pmatrix}  1 & &  \\  & 0 & \\  & & \ddots \end{pmatrix} + \ldots + \lvert c_d \rvert ^2 \begin{pmatrix}  \ddots & &  \\  & 0 & \\ & & 1 \end{pmatrix}$, which is the state obtained after decoherence entailed by the measurement apparatus (non-selective measurement \eqref{non-selective}), to the collapsed state $\begin{pmatrix}  1 & &  \\  & 0 & \\  & & \ddots \end{pmatrix}$ (selective measurement \eqref{selective}).

However, this formulation of the problem makes little sense; troubles appear when one believe more in the mathematics than in the world they are trying to describe. Of course, no observer has ever observed more than one fact at once, and no one has ever felt superposed. Also, one must not demand more to QM than it can give: 
\begin{itemize}
\item being a probabilistic theory, it is normal that several possibilities appear, and that only an experimental observation can determine the actual result
\item being a probabilistic theory, it is also very natural that an update has to be performed when one obtains new empirical information about the system. This is a common feature of any probabilistic model\footnote{As a trivial example, suppose for instance that you ask your computer to choose uniformly a number $x_1$ in $\{0 ; 1\}$, and then to choose uniformly a second number $x_2$ in $\{0 ; 1\}$ if $x_1=0$, or to choose uniformly $x_2$ in $\{2 ; 3\}$ if $x_1=0$. If you don't look at the result for $x_1$, you predict $x_2$ to follow a uniform law in $\{0 ; 1; 2 ; 3\}$, but if you know the outcome for $x_1$, you obviously update your probabilities. \label{trivial_example}}.
\end{itemize}

Of course, in QM, things are more confusing because it is a probabilistic theory in which all the possible histories interfere together, so that every potential outcome seems to have had an influence on the result of the experiment, hence a sort of ‘reality’. This is why the collapse looks so physical (at least more than a simple subjective information update as in the example of note \ref{trivial_example}), so that the existence of a truly physical process leading from $\begin{pmatrix}  \lvert c_1 \rvert ^2 & &  \\  & \ddots & \\  & &  \lvert c_d \rvert ^2 \end{pmatrix}$ to the observed  $\begin{pmatrix}  1 & &  \\  & 0 & \\  & & \ddots \end{pmatrix}$ has been proposed. This is the starting point for the research on objective collapse models, trying to add a non-linear stochastic modification of the Schrödinger equation which would explain the transition from mixed to deterministic states at macroscopic scales \citep{bassi2013models}. In this case, the question at stake is roughly: when exactly does the collapse occur, and at which speed? However, while those models suffer from issues in their compatibility with special relativity despite numerous attempts \citep{dove1996explicit, tumulka2006relativistic,bedingham2014matter}, large portions of the set of possible parameters have already been ruled out experimentally \cite{carlesso2022present}. They are also at odds with the phenomenon of coherence revival (see \S\ref{physical_collapse}).

On the other hand, if the collapse is not a physical process but a mere epistemic update as in note \ref{trivial_example}, it remains to understand why it still yields correct predictions. Indeed, the operation of updating the probabilities by suppressing all the possible histories except the one that has actually been observed (or the ones compatible with the observation) is legitimate as long as re-coherence is not permitted, \textit{i.e.} as long as $\eta$ remains 0 forever. Otherwise, the possible histories interfere again so that suppressing some of them should lead to wrong predictions (as in the experiment of \S\ref{physical_collapse}). Said differently, in the case of a non-physical collapse, determining exactly when does the latter occur is no more problematic (you update your mathematical tool whenever you wish), but one has to understand what always justifies its use (see \S\ref{definition_measurement}).

\subsection{Is the collapse a physical process?} \label{physical_collapse}

In \cite{bassi2013models}, it is claimed that many-worlds interpretations based on decoherence can not be experimentally distinguished from the Copenhagen interpretation. But this is not correct since no interpretation assuming the reality of the collapse (Copenhagen or collapse models) can explain a phenomenon such as coherence revival, as sketched in the following thought experiment. Note first that when a quantum system like a spin $\frac{1}{2}$ interacts with a microscopic environment like another spin $\frac{1}{2}$, decoherence is possible but re-coherence is not difficult to achieve, for example \textit{via} a C-NOT gate (by default, the spin is considered along the $z$ axis): 

\[  \underbrace{\frac{1}{\sqrt{2}} (\ket{\uparrow} + \ket{\downarrow}) }_{\mathcal{S}} \otimes \underbrace{ \ket{\uparrow} }_{\mathcal{E}} \underset{C-NOT} \longrightarrow \underbrace{ \frac{1}{\sqrt{2}} (\ket{\uparrow \uparrow} + \ket{\downarrow \downarrow}) }_{\rho_{\mathcal{S}} =  \begin{pmatrix} \frac{1}{2} & 0 \\ 0 & \frac{1}{2} \end{pmatrix} } \underset{C-NOT} \longrightarrow \frac{1}{\sqrt{2}} (\ket{\uparrow} + \ket{\downarrow}) \otimes \ket{\uparrow}. \]
Here $\mathcal{S}$ clearly decoheres and then retrieves its coherence at the end in the superposed state $\frac{1}{\sqrt{2}} (\ket{\uparrow} + \ket{\downarrow}) = \ket{\uparrow_x}$, for which a measurement of the spin along $x$ yields $\uparrow_x$ with probability 1. The same operation could, in principle, be applied to a system coupled with a much larger environment, modeled for example by a collection of $n$ spins with huge $n$. Note that it doesn't have to be the genuine spin, but could also stand for any abstract degree of freedom, in particular it could be linked with position which is generally the basis in which the collapse is induced in collapse models. If the collapse had a physical reality, coupling $\mathcal{S}$ with a macroscopic system should make it undergo the collapse \eqref{selective}, that will affect the statistics after re-coherence: 

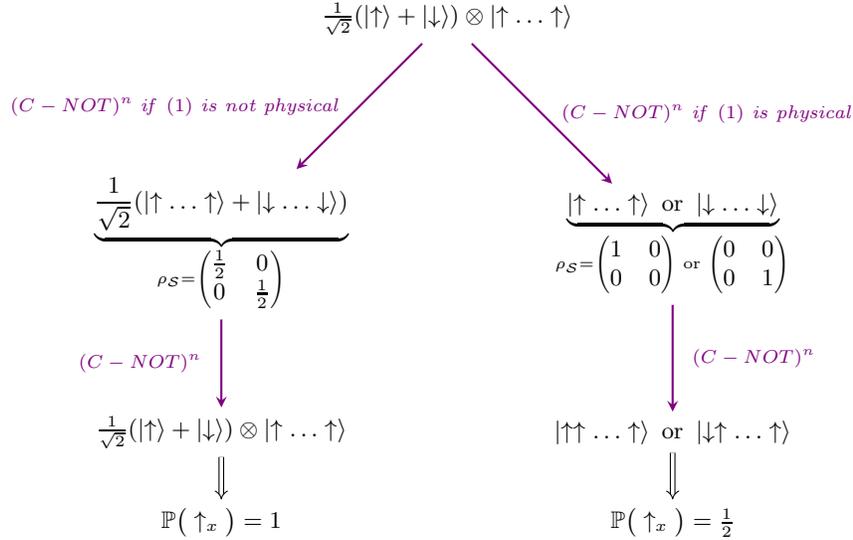
\begin{center}
\begin{tikzpicture}[node distance = 3cm,auto]
\node (1) {$\frac{1}{\sqrt{2}} (\ket{\uparrow} + \ket{\downarrow})  \otimes \ket{\uparrow \dots \uparrow}$};
\node [below of = 1] (void) {};
\node [right of = void] (2) {$\underbrace{ \ket{\uparrow \dots \uparrow} \text{ or } \ket{\downarrow \dots \downarrow} }_{\rho_{\mathcal{S}} =  \begin{pmatrix} 1 & 0 \\ 0 & 0 \end{pmatrix} \text{ or } \begin{pmatrix} 0 & 0 \\ 0 & 1 \end{pmatrix}}$ } ;
\node [left of = void] (3) {$\underbrace{ \frac{1}{\sqrt{2}} (\ket{\uparrow \dots \uparrow} + \ket{\downarrow \dots \downarrow})}_{\rho_{\mathcal{S}} =  \begin{pmatrix} \frac{1}{2} & 0 \\ 0 & \frac{1}{2} \end{pmatrix}}$ };
\node [below of = 2, node distance=2.5cm] (4) {$\ket{\uparrow \uparrow \dots \uparrow} \text{ or } \ket{\downarrow \uparrow \dots \uparrow}$};
\node [below of = 3, node distance=2.5cm] (5) {$\frac{1}{\sqrt{2}} (\ket{\uparrow} + \ket{\downarrow})  \otimes \ket{\uparrow \dots \uparrow}$};
\node [below of = 4, node distance=1.2cm] (6) {$\mathbb{P}\big( \uparrow_x \big) = \frac{1}{2}$};
\node [below of = 5, node distance=1.2cm] (7) {$\mathbb{P}\big( \uparrow_x \big) = 1$};
\draw [thick, -stealth, violet] (1) -- (2) node[midway, right=4] {\color{violet} \scriptsize $(C-NOT)^n$ \textit{if \eqref{selective} is physical}};
\draw [thick, -stealth, violet] (1) -- (3) node[midway, left=4] {\color{violet} \scriptsize $(C-NOT)^n$ \textit{if \eqref{selective} is not physical}};
\draw [thick, -stealth, violet] (2) -- (4) node[midway, right=4] {\color{violet} \scriptsize $(C-NOT)^n$} ;
\draw [thick, -stealth, violet] (3) -- (5) node[midway, left=4] {\color{violet} \scriptsize $(C-NOT)^n$};
\draw  (4)  edge [-implies,double equal sign distance]  (6);
\draw  (5)  edge [-implies,double equal sign distance]  (7);
\end{tikzpicture}
\end{center} 

Obviously, this experiment is not feasible in the present technology of quantum computing for a large $n$, but more realistic ones have been conducted and coherence revival is already a confirmed phenomenon \citep{chapman1995photon}. At the end of the day, we still have no compelling empirical reason to believe that a collapse process exists whereas, as recalled above, an update operation is naturally expected in any probabilistic theory. Moreover, for the first time since the advent of QM, the theory of decoherence allows us to assert consistently that the Schrödinger equation is universally valid at all scales (see \S\ref{second_formulation}). What can be the status of decoherence in a collapse model, apart from a curious redundant phenomenon that also happens to destroy the quantum interferences? In the following, we will therefore consider that the collapse is not a physical process; it remains to address the problem stated above at the end of \S\ref{epistemic/ontic}.

\subsection{Definition of the measurement} \label{definition_measurement}

By chance, as explained by the theory of decoherence, it is impossible for an observer to obtain empirical information on a quantum system without (almost) immediately and perfectly destroying the quantum interferences. Said differently, a knowledge update (selective measurement) \eqref{selective} is inevitably preceded by the physical process (non-selective measurement) \eqref{non-selective}\footnote{If it were not the case, the universe would be very different... This is also why it is so hard to untangle the two and clarify their respective status.}. As mentioned above, as long as coherence is not retrieved, the possible histories will never interfere anymore, so that forgetting about those which did not happen will not change any future prediction. It is important to keep in mind, though, that this ideal case is never absolutely fulfilled, because decoherence is not immediate nor perfect, and because re-coherence can happen in particular experiments (as in the example of \S\ref{physical_collapse}). It is even \textit{expected} in principle, even though the recurrence time can easily exceed the lifetime of the universe for realistic systems \citep{zurek1982environment}. 

However, the collapse follows a gain of information due to a measurement. In this case, the situation is much better because the physicist who reads, \textit{remembers} and stores the data of the outcome is part of the environment too. As long as these data exist materially\footnote{By ‘matter’, we designate all that can be objectively measured and quantified.} (be it in her brain, her computer or her notebook), the vectors $(\ket{\mathcal{E}_i(t)})_{1\leq i \leq d}$ are perfectly distinguishable, hence orthogonal, guaranteeing $\eta$ to be 0. Thus re-coherence of the possible histories can only happen when nothing material remains that allows to know what the outcome was. In other words, when a measurement has been performed, a condition for the collapse to ultimately lead to wrong predictions is that nothing in the universe should remember the outcome\ldots\ but in this case the person who made the prediction is not here anymore willing to check it! (More on this in \S\ref{consistency}.) \\

To build an interpretation of a theory is all about finding a way to present the mathematics, and developing a vocabulary to speak of those mathematics, that raise clear pictures in mind with as few counter-intuitive properties as possible. In \S\ref{remarks on QM}, we explained why quantum theory is so reluctant to ontological pictures. Nevertheless, this is not a reason for completely neglecting the search for a (less worse) interpretation. In fact, a good language brings good reasonings and, all things considered, (interpretation of the) theory and experiments co-evolve, as brilliantly pointed by Einstein in this discussion with Heisenberg, related in Heisenberg's autobiographical notes \citep{heisenberg1973development}: ‘It may be of heuristic value to recall what one really observes. But from a principal point of view it is quite wrong to insist on founding a theory on observed quantities alone. In reality just the opposite is true. Only the theory decides about what can be observed.’ It is indeed the interpretation of the theory that dictates the langage with which we describe our observations, the meaning we give to them, and also influences the instruments we build and the experiments we decide to conduct. In particular, the usual expression ‘the particle is both here and there’ generally used to refer to a superposition \textit{is} an interpretation, and arguably a quite confusing choice of words. Let's try to introduce some new vocabulary.

\begin{definition}[\textbf{Measurement in quantum mechanics}] \label{definition}
\text{ } \begin{itemize}

\item We say that the observable $A$ of a system $\mathcal{S}$ is \textbf{recorded} by an environment $\mathcal{E}$, when $\mathcal{S}$ and $\mathcal{E}$ are entangled with perfect decoherence in the eigenbasis $\mathcal{B}$ of $\hat{A}$ (\textit{i.e.} $\eta_{\mathcal{B}}=0$). As long as this holds, we say that the possible histories of $\mathcal{S}$ are \textbf{split} by $\mathcal{E}$ in $\mathcal{B}$, or simply that $A$ is split. If coherence is retrieved, we say that the possible histories \textbf{recombine}.

\item A \textbf{non-selective measurement process} for $A$ is a unitary evolution that entangles $\mathcal{S}$ with $\mathcal{E}$ such that, \textit{for any initial state} of $\mathcal{S}$, the system's possible histories are split in $\mathcal{B}$ (\textit{i.e.} $A$ is recorded by $\mathcal{E}$) after interaction. If moreover $\mathcal{B}$ is a conserved basis during the process, we call it a \textbf{projective} (or non-demolition) measurement. 

\item A \textbf{selective measurement} of $A$ by an observer $\mathcal{O}$ occurs when $A$ is split (in particular) by the act of $\mathcal{O}$ getting entangled with $\mathcal{S}$ when storing the data of an outcome. As long as this holds, we say that the observed outcome is a \textbf{fact} for $\mathcal{O}$ and that $\mathcal{O}$ \textbf{remembers} the data of the outcome. The \textbf{collapse} is the subsequent update performed by $\mathcal{O}$ on the probabilities, on the basis of this new knowledge.

\end{itemize}
\end{definition} 

The first is only an instantaneous statement: does the environment presently distinguishes the eigenstates of $A$? Mathematically, this means that $\mathcal{S}+\mathcal{E}$'s state takes the form $\sum_i c_i \ket{A_i} \ket{\mathcal{E}_i}$, where $\mathcal{B} = (\ket{A_i})_i$ are the eigenstates of $A$ and $\braket{\mathcal{E}_i \vert \mathcal{E}_j} = 0$ for $i \neq j$. In this case, the total probability formula applies, so that one can assume that we are in one — and only one — of the possible histories concerning $A$. Note that any entanglement constitutes a recording in the basis where the density matrix of $\mathcal{S}$ is diagonal after interaction\footnote{A slight redefinition of $\eta_{\mathcal{B}}$ as $\displaystyle \max_{\substack{i \neq j \\ c_i, c_j \neq 0}} \; \lvert \braket{\mathcal{E}_i(t) \vert \mathcal{E}_j(t)} \rvert$ with, by convention, $\eta_{\mathcal{B}} = 0$ when only one $c_i$ is non zero, may be necessary to ensure that $\rho_{\mathcal{S}}$ being diagonal in $\mathcal{B}$ is equivalent to $\eta_{\mathcal{B}}=0$. Further adaptations could be made to deal with the possible degeneracy of $\hat{A}$'s eigenvalues, which need not necessarily be distinguished by the environment\dots} (this basis may depend on the initial state of the system and/or of the environment).

The second, in the projective case, is a condition for an interaction to entail the operation \eqref{non-selective}. It relies on the existence of a unitary $U$ and an initial environment state $\ket{\mathcal{E}_0}$ such that, for all $i$, $U \ket{A_i} \ket{\mathcal{E}_0} = \ket{A_i} \ket{\mathcal{E}_i}$ with $\braket{\mathcal{E}_i \vert \mathcal{E}_j} = 0$ for $i \neq j$. But the existence of a conserved basis is a strong requirement, and we consider here that it is enough to have $\mathcal{E}$ split the possible histories in a given basis in order to speak of measurement (only a projective measurement reflects something of the initial state; it also differs from a POVM, in which the environment distinguishes what the system's state was \textit{before} interaction, no matter if its final state is altered). In this case, we only suppose that for all $\ket{\Psi} \in \mathcal{H}_{\mathcal{S}}$, $U \ket{\Psi} \ket{\mathcal{E}_0} = \sum_i c_i^\Psi \ket{A_i} \ket{\mathcal{E}^\Psi_i}$ with $\braket{\mathcal{E}^\Psi_i \vert \mathcal{E}^\Psi_j} = 0$ for $i \neq j$. By abuse of language when the context is clear, we may still write ‘measurement’ instead of ‘projective measurement’ as we did above.

The third is a condition for $\mathcal{O}$ to apply safely the collapse \eqref{selective} and get correct predictions, at least as long as she remembers the observed outcome. It is based on the fact that obtaining some knowledge about a system necessarily splits its possible histories in the basis corresponding to the question asked, because $\mathcal{S}+\mathcal{O}$'s state takes the form $\sum_i \ket{A_i} \ket{\mathcal{O}_i(t)}$, where $\braket{\mathcal{O}_i(t) \vert \mathcal{O}_j(t)} = 0$ for $i \neq j$ as long as $\mathcal{O}$ remembers the outcome. Since we consider that the collapse is not a physical operation, it is natural that its definition doesn't entirely rely on physical concepts. By observer, we designate at least any human being that can remember facts; the question of what exactly can be deemed an observer is addressed in \S\ref{observer}. We see that it is not wrong, for example, to say that the primordial universe has collapsed the first time an astrophysicist has looked at the cosmic microwave background, simply because it was the first time that someone acquired knowledge about the latter, although the possible histories of each primordial photon were split long ago by decoherence. We also see that there is one wave function per observer, as any probability distribution depends on the information available to the person making predictions\footnote{Mathematics have long ago accepted the idea that a theory is a point of view, always embedded in a larger meta-theory, itself subject to reasonings applicable to theories. There is no more absolute wave function of the universe than there is an ultimate mathematical theory.}. However, due to the particular way of computing probabilities in QM, it must be checked that no inconsistencies appear when observers who have different knowledge compare their predictions (see \S\ref{consistency}). 

\section{Second formulation: the transition from quantum to classical physics} \label{second_formulation}
The other main aspect of the measurement problem concerns the emergence of classicality. If no physical collapse is assumed, a myriad a interrogations must be solved with regard to the seemingly extremely different aspects of quantum and classical physics. Can the Schrödinger equation really be considered valid at all scales?

\subsection{Some remarks on Schrödinger's cat} \label{Schrödinger's cat}

Schrödinger's cat paradox is often cited as a key illustration of the measurement problem. The basic question is: why don't we observe cats in superpositions like $\alpha \ket{\text{dead}} + \beta \ket{\text{alive}}$\ ? Once again, this quick formulation is too simplistic for the reasons listed below, but the thought experiment still contains interesting issues concerning the applicability of QM to our daily environment.

\begin{itemize}
\item Even if the room were perfectly isolated, the cat would already be completely decohered by the Geiger counter composing its environment, or even by the unstable atom only, hence in a mixed state with no quantum interferences. In the following, let's rather suppose that the cat is truly isolated.

\item The level of interferences is in general basis-dependant, so even if the cat is not superposed in the basis $\mathcal{B} = (\ket{\text{dead}} , \ket{\text{alive}})$, it still presents quantum interferences in the vast majority of other basis. What is so special with $\mathcal{B}$ from our point of view and, to begin with, of which Hilbert space is it a basis (see \S\ref{preferred_basis})?   

\item How could we possibly know whether the cat is superposed or not? Clearly not by opening the room and looking at it, which would constitute a measurement in the basis $\mathcal{B}$, thus immediately destroying the interferences between the histories $\ket{\text{dead}}$ and $\ket{\text{alive}}$. As mentioned earlier, a cat in superposition should preferably not been thought of as being both dead and alive, but simply as a cat whose statistics are not those of a classical cat either dead or alive. To empirically reveal the presence of a superposition, one has to make the histories interfere, for example by performing a measurement in another orthogonal basis like $\mathcal{B}' = (\alpha \ket{\text{dead}} + \beta \ket{\text{alive}} , \bar{\beta} \ket{\text{dead}} - \bar{\alpha} \ket{\text{alive}})$ or, equivalently, to make the cat undergo a unitary evolution sending $\mathcal{B}'$ to $\mathcal{B}$ before performing a measurement in $\mathcal{B}$, and repeat this process several times to check that we indeed obtain $\alpha \ket{\text{dead}} + \beta \ket{\text{alive}}$ in the first case, or $\ket{\text{dead}}$ in the second case, with probability 1. But what kind of apparatus could well perform a measurement with respect to $\mathcal{B}'$\ ? How could we have such absolute quantum control on the cat so that we can rotate its state from $\mathcal{B}'$ to $\mathcal{B}$\ ? Furthermore, how could we prepare a large number of cats in exactly the same quantum state? The problem is that many aspects of \textit{QM are not even testable at our scales}, so we can't do better than to check its compatibility, in principle, with what we know of classical physics. As we will see throughout this section, we have no reason to think that quantum theory is not universally valid because, if it is, then the oddities that \textit{could} appear at our scales are actually prevented by the fact that we can not play God. In particular, we can neither perfectly isolate a large system nor apply any unitary evolution we wish (this is why we never observe recombination of macroscopic histories as in the thought experiment 2.\ of \S\ref{consistency}), nor prepare a precise macroscopic state, and we only have access to a very small number of coarse measurements (see \S\ref{preferred_basis}).
\end{itemize}

\subsection{The preferred basis problem} \label{preferred_basis}

At this point, a comment must be made on the status of what we call the possible histories. In QM, choosing an orthonormal basis $\mathcal{B}$ of the Hilbert space corresponds to choosing a particular way to tell the possible histories, and the value of $\eta_\mathcal{B}$ in a given basis quantifies to what extent these histories are split. We have $\eta_{\mathcal{B}} = 0$ only in an eigenbasis of $\rho_{\mathcal{S}}$, so decoherence only ensures that the histories are easy to tell in this particular basis where they don't influence each other. The very fact for a system to have its possible histories split or interfere (some might say: for worlds to branch), is not absolute because there are infinitely inequivalent ways to decompose them. The picture of a well-defined graph of histories splitting over time with decoherence, and recombining with coherence revival, is a bit too simple. 

Nonetheless, there exists cases in which all the histories are split in any basis. Consider for instance a spin $\frac{1}{2}$ entangled with a measurement apparatus $\mathcal{A}$ in the state $\ket{\Psi} = \frac{1}{\sqrt{2}} \big( \ket{\uparrow \mathcal{A}_{\uparrow}} + \ket{\downarrow \mathcal{A}_{\downarrow}} \big)$, such that the spin along $z$ is split by $\mathcal{A}$ \textit{i.e.} $\braket{\mathcal{A}_{\uparrow} \vert  \mathcal{A}_{\downarrow}} = 0$. It is not difficult to show that one can rewrite, for any other direction $u$, $\ket{\Psi} = \frac{1}{\sqrt{2}} \left( \ket{\uparrow_u \mathcal{A}_{\uparrow_u}} + \ket{\downarrow_u \mathcal{A}_{\downarrow_u}}  \right)$, with $\ket{\mathcal{A}_{\uparrow_u}}$ and $\ket{\mathcal{A}_{\downarrow_u}}$ some linear combinations of $\ket{\mathcal{A}_{\uparrow}}$ and $\ket{\mathcal{A}_{\downarrow}}$ still satisfying $\braket{\mathcal{A}_{\uparrow_u} \vert  \mathcal{A}_{\downarrow_u}} = 0$. The reason is that all the possible histories are equiprobable so the density matrix is a scalar matrix which is diagonal in every basis. Therefore, if being split by an environment were a sufficient criterion to define the selective measurement, it would be possible here to apply the collapse in \textit{any} basis, which is manifestly wrong! As stated by \cite{schlosshauer2005decoherence}, QM ‘has nothing to say about which observable(s) of the system is (are) being recorded, via the formation of quantum correlations, by the apparatus (...) in obvious contrast to our experience of the workings of measuring devices that seem to be “designed” to measure certain quantities’. 

So what can well distinguish the basis in which the spin was really measured? Here, if ever $\ket{\mathcal{A}_{\uparrow}}$ is the state of a device displaying a definite outcome, $\ket{\mathcal{A}_{\uparrow_u}}$ is not. Indeed, a measurement apparatus can be seen as a tool associating the eigenvectors of the observable to be measured in $\mathcal{H}_\mathcal{S}$ with (in a good approximation) eigenvectors of the position observable in $\mathcal{H}_\mathcal{A}$. In this way, reading the outcome is easily accessible to us; in fact, we wouldn't know how to do otherwise anyway: no engineer has ever wondered ‘Into which orthogonal vectors do I want my device to evolve when measuring a spin purely up or down?’. But if $\ket{\mathcal{A}_{\uparrow}}$ is spatially localized, all the other $\ket{\mathcal{A}_{\uparrow_u}}$ are spatial superpositions. This characterizes the basis in which the measurement took place. 

The position eigenbasis clearly plays a special role in QM (see \cite{bell1982impossible}), especially when the system goes larger. It is generally a preferred basis for decoherence \citep{schlosshauer2005decoherence}, certainly because the laws of physics usually involve position variables, hence environments accurately feel variations in position and record them. For the same reason it is also the basis in which we can most easily perform a measurement, be it very imprecise, at our scales (sending a few photons may suffice to roughly localize an object). Said differently, the salient information we have about the world is mostly linked with position, so this is also in this basis that we conceive our interaction with it, as does the engineer cited above\footnote{This is actually a deep remark, but one should be very careful with its philosophical implications. I reckon that it would not be correct to state that the human body and mind (and presumably all living beings on Earth) evolved so as to develop perceptions strongly based on the notion of space and position because this evolutionary path allowed them to benefit from a stability of facts entailed by decoherence on this observable. Indeed, there was no ‘notion of space and position’ before something developed it, and even less any ‘laws of physics involving position variables’, since such laws only belong to the worlds of those sensitive beings whose minds are already appropriately shaped. I would be surprised to meet an alien, but I would not be surprised if it has absolutely nothing like a ‘notion of space and position’ even though its body, if it is made of matter, will certainly be perceived by us as obeying the laws of physics. This alien will probably not survive long on Earth, but maybe on its planet there is no heredity and therefore no Darwinian evolution; maybe the stability of facts, or the very notion of facts, don't confer there any kind of advantage; maybe what we would deem to be its death would not be a relevant event in its world. If ever it has something like a physics, I would try with all my energy to understand it. When talking about the world from outside the earthly point of view, there is really nothing that can be expected. The earthly minds and bodies on one side, their world of outer perceptions on the other side, co-evolved; it would be misleading but appreciably poetic to say that they emerged out of the amorphous void. \label{note_philo}}. One may speculate that the position basis may be defined as the basis in which measurements require the least amount of energy for an observer (about measurements perturbing the system, see \S\ref{determinism} and \S\ref{consistency}).

Even on particles, we are able to measure only a very restricted set of definite observables: position, momentum, spin, charge, energy, angular momentum... But on macroscopic systems, we have much less: we are restricted to very coarse measurements that discriminate only between huge subspaces of $L^2(\mathbb{R}^3)$, which are typically unions of eigenspaces of the position observable. For instance, measuring the cat in the basis $(\ket{\text{dead}} , \ket{\text{alive}})$ precisely means that, by looking at it (\textit{i.e.} exchanging some photons), we entangle with it and split two large sets of its possible histories, which we term the ‘dead’ and ‘alive’ subspaces. These latter are much more related to the position operator (‘is the cat standing or lying on the floor?’) than any other observable (we don't care about questions like ‘what is the internal energy of the cat?’). 

This is the grand lesson of the theory of decoherence: even though QM is universally valid, our experience of the world is governed by classical probabilities because the structure of the laws of physics make position a preferred basis for splitting histories, and this holds for us as well as for any environment, therefore the observables we are mainly able to measure on a system are rightly the same as those continuously recorded by its environment. We only distinguish between histories of a system that are already split by the rest of the world.

\subsection{The disappearance of quantum correlations} \label{correlations}

Any physical interaction creates entanglement. If everything is so entangled, why don't we experience its purely quantum effects at our scales, namely correlations violating Bell's inequality? In a Bell experiment, one well-chosen measurement is performed on each particle of a Bell state $\ket{\Psi} = c_1 \ket{\uparrow \downarrow} + c_2 \ket{\downarrow \uparrow}$. For macroscopic entangled pairs, however, none of the two elements are in practice isolated. They have been separately recorded by their respective environments $\mathcal{E}$ and $\mathcal{E}'$ in some bases $\mathcal{B}$ and $\mathcal{B}'$ (namely in the eigenbases of their density matrices after interaction). The two interactions taking place in different Hilbert spaces, they are described by commuting Hamiltonians so we can deal with them one after the other. Let's still take $c_1 \ket{\uparrow \downarrow} + c_2 \ket{\downarrow \uparrow}$ as the initial state of the pair, $\ket{\mathcal{E}_0}$ and $\ket{\mathcal{E}'_0}$ the initial states of the environments. Denote $\mathcal{B} = ( \alpha \ket{\uparrow} + \beta \ket{\downarrow} , \bar{\beta} \ket{\uparrow} - \bar{\alpha} \ket{\downarrow}) \equiv (\ket{\uparrow}_{\mathcal{B}} , \ket{\downarrow}_{\mathcal{B}})$ and $\mathcal{B}' = ( \alpha' \ket{\uparrow} + \beta' \ket{\downarrow} , \bar{\beta}' \ket{\uparrow} - \bar{\alpha}' \ket{\downarrow}) \equiv (\ket{\uparrow}_{\mathcal{B}'} , \ket{\downarrow}_{\mathcal{B}'})$. For simplicity, we will assume that $\mathcal{B}$ and $\mathcal{B}'$ are conserved bases, so that the interactions are simply defined by: $\ket{\mathcal{E}_0} \ket{\uparrow}_{\mathcal{B}} \rightsquigarrow \ket{\mathcal{E}_{\uparrow_{\mathcal{B}}}} \ket{\uparrow}_{\mathcal{B}}$ ; $\ket{\mathcal{E}_0} \ket{\downarrow}_{\mathcal{B}} \rightsquigarrow \ket{\mathcal{E}_{\downarrow_{\mathcal{B}}}} \ket{\downarrow}_{\mathcal{B}}$ with $\braket{\mathcal{E}_{\uparrow_{\mathcal{B}}} \vert \mathcal{E}_{\downarrow_{\mathcal{B}}}}= 0$ by construction, and similarly with the primes. Here is what happens schematically:

\begin{align*}
& \ket{\mathcal{E}_0} \Big(c_1 \ket{\uparrow \downarrow} + c_2 \ket{\downarrow \uparrow} \Big) \ket{\mathcal{E}'_0} \\
& \quad = \ket{\mathcal{E}_0} \Big(c_1 (\bar{\alpha} \ket{\uparrow}_{\mathcal{B}} + \beta \ket{\downarrow}_{\mathcal{B}}) \otimes \ket{\downarrow} + c_2 (\bar{\beta} \ket{\uparrow}_{\mathcal{B}} - \alpha \ket{\downarrow}_{\mathcal{B}}) \otimes \ket{\uparrow} \Big) \ket{\mathcal{E}'_0} \\ \\
\underset{\mathcal{E} \text{ records}}{\rightsquigarrow} \quad & \ket{\mathcal{E}_{\uparrow_{\mathcal{B}}}}  \ket{\uparrow}_{\mathcal{B}} \Big(c_1 \bar{\alpha} \ket{\downarrow} + c_2  \bar{\beta} \ket{\uparrow} \Big) \ket{\mathcal{E}'_0} +  \ket{\mathcal{E}_{\downarrow_{\mathcal{B}}}}  \ket{\downarrow}_{\mathcal{B}} \Big(c_1 \beta \ket{\downarrow} - c_2 \alpha \ket{\uparrow}\Big)  \ket{\mathcal{E}'_0}  \\
& \quad = \ket{\mathcal{E}_{\uparrow_{\mathcal{B}}}}  \ket{\uparrow}_{\mathcal{B}} \Big(c_1 \bar{\alpha}  [ \bar{\beta}' \ket{\uparrow}_{\mathcal{B}'} - \alpha' \ket{\downarrow}_{\mathcal{B}'}] + c_2  \bar{\beta}  [ \bar{\alpha}' \ket{\uparrow}_{\mathcal{B}'} + \beta' \ket{\downarrow}_{\mathcal{B}'}]  \Big) \ket{\mathcal{E}'_0} \\
& \quad \quad +  \ket{\mathcal{E}_{\downarrow_{\mathcal{B}}}}  \ket{\downarrow}_{\mathcal{B}} \Big(c_1 \beta [ \bar{\beta}' \ket{\uparrow}_{\mathcal{B}'} - \alpha' \ket{\downarrow}_{\mathcal{B}'}] - c_2 \alpha [ \bar{\alpha}' \ket{\uparrow}_{\mathcal{B}'} + \beta' \ket{\downarrow}_{\mathcal{B}'}]  \Big)  \ket{\mathcal{E}'_0} \\ \\
\underset{\mathcal{E}' \text{ records}}{\rightsquigarrow} \quad & (c_1 \bar{\alpha} \bar{\beta}' +  c_2  \bar{\beta} \bar{\alpha}' )  \ket{\mathcal{E}_{\uparrow_{\mathcal{B}}}}  \ket{\uparrow}_{\mathcal{B}} \ket{\uparrow}_{\mathcal{B}'}  \ket{\mathcal{E}'_{\uparrow_{\mathcal{B}'}}} + (-c_1 \bar{\alpha} \alpha' + c_2 \bar{\beta} \beta') \ket{\mathcal{E}_{\uparrow_{\mathcal{B}}}}  \ket{\uparrow}_{\mathcal{B}} \ket{\downarrow}_{\mathcal{B}'}  \ket{\mathcal{E}'_{\downarrow_{\mathcal{B}'}}} \\
& + (c_1 \beta \bar{\beta}' - c_2 \alpha \bar{\alpha}') \ket{\mathcal{E}_{\downarrow_{\mathcal{B}}}}  \ket{\downarrow}_{\mathcal{B}} \ket{\uparrow}_{\mathcal{B}'}  \ket{\mathcal{E}'_{\uparrow_{\mathcal{B}'}}} + (-c_1 \beta \alpha' - c_2 \alpha \beta') \ket{\mathcal{E}_{\downarrow_{\mathcal{B}}}}  \ket{\downarrow}_{\mathcal{B}} \ket{\downarrow}_{\mathcal{B}'}  \ket{\mathcal{E}'_{\downarrow_{\mathcal{B}'}}}
\end{align*}
The pair's possible histories are now split in $\mathcal{B} \otimes \mathcal{B}'$ (because $\braket{\mathcal{E}_{\uparrow_{\mathcal{B}}} \vert \mathcal{E}_{\downarrow_{\mathcal{B}}}}= \braket{\mathcal{E}_{\uparrow_{\mathcal{B}'}} \vert \mathcal{E}_{\downarrow_{\mathcal{B}'}}} = 0$). Even if we don't know what the environments have actually recorded, the total probability formula applies and we can assert that we are in only one of these histories. The important thing is that \textit{they all take the form of pure tensor products}: no violation of Bell's inequalities can occur with such non-entangled systems. Note that it doesn't matter if $\mathcal{B}$ and $\mathcal{B}'$ are changing over time; the environments don't have to record the value of any fixed observable. 

There are, however, two difficulties with this proof:
\begin{itemize}
\item the above proof does not work in general if the bases are not assumed conserved anymore,
\item we supposed the two environments initially non-entangled; a justification could be given by the fact that they are themselves already recorded by the rest of the world, but one should be careful to avoid circular arguments\dots
\end{itemize}

\subsection{The emergence of determinism} \label{determinism}

An issue that is not often related to the measurement problem, and yet has much to do with measurements, is the following question: if QM is valid at all scales, hence if classical physics stems from QM, how can the former be deterministic while the latter is probabilistic? Why is our experience of the world so predictable in a quantum universe?

QM, as it stands, can not provide fully deterministic predictions in particular because of Heisenberg inequality, which is a theorem implied by the postulates of the theory. Presented this way, one might think that QM is simply incomplete and that, one day, we could have at our disposal a better theory predicting, for each single experiment, the exact outcome (a hidden variable theory). However, we should recall that Heisenberg initially justified his ‘principle’ with various heuristic arguments before it became a theorem. Most people seem to agree that the theorem formulation is more satisfying and much stronger but, all things considered, it is quite the contrary. The heuristic reasonings are based on the universal fact that \textit{measuring a system requires to exchange energy with it, and therefore perturbs it} (item 2.\ (ii) in \S\ref{remarks on QM}): this fondamental idea might really be called the Heisenberg uncertainty principle, and is no more present in the theorem (the perfect agreement between the principle and the theorem is actually very puzzling, so their status, content and rigidity differ). The consequence of this principle is the impossibility to access exact initial conditions for particles, which is a necessary condition for any theory to provide deterministic predictions. It is valid independently of the theory and this is precisely the reason why, even though we already do have a hidden variable theory compatible with QM, namely Bohmian mechanics, the latter is exactly as uncertain as QM.

Why, then, does this principle doesn't affect classical physics? There are at least two main reasons for that.
\begin{itemize}
\item Until the 19th century, we hadn't empirical access to systems whose typical action is of order $\hbar$, for which a measurement entails a significant uncertainty on the initial condition itself, hence immediate indeterminism. Of course, uncertainties propagate in time and may grow dramatically. But for systems like the solar system for instance, the uncertainty on the initial condition we have (much larger than $\hbar$!) remains sufficiently long under the uncertainty of our own measurement apparatus, and so is practically undetectable.
\item Classical physics also deals with systems that are chaotic (\textit{i.e.} with large Lyapunov exponents, as does statistical physics) or composed of particles subject to a notable uncertainty (like photons in optics and electromagnetism). In these cases, we restrict ourselves to a well-chosen variables that become practically deterministic thanks to the law of large numbers.
\end{itemize}
One last word about the largely debated question of whether the brain has quantum properties or should be treated as a classical system. Maybe this is finally of little interest compared to the question of knowing whether the brain can be considered as deterministic \textit{i.e.} fully predictable: given the best initial condition we can hope for a brain, how long will the indeterminacy remain negligible?

\subsection{Checking the consistency of quantum mechanics at all scales} \label{consistency}

We now understand why the peculiar quantum features (interferences, quantum correlations, indeterminism) are not observed at our scales. From a theoretical point of view, though, we should also make sure that the assumption of the universal validity of QM has no internal inconsistencies, even for omnipotent observers who could have absolute quantum control on arbitrary systems. Potential contradictions could at first sight occur when different observers having different knowledge, hence using different wave functions, compare their predictions (as evoked in \S\ref{definition_measurement}). This situation is captured by the famous Wigner's friend thought experiment. In what follows, we will grant more and more quantum powers to Wigner and see if he and his friend expect the same statistics for the outcomes of their joint experiment. 

\begin{enumerate}
\item Let him first be able to perfectly isolate together a spin $\frac{1}{2}$ and his friend $\mathcal{F}$ measuring it along $z$, and then isolate the particle from the friend to perform a measurement on the spin along a direction $u$ characterized by $\ket{\uparrow} = \alpha \ket{\uparrow_u} + \beta \ket{\downarrow_u}$ and $\ket{\downarrow} = \bar{\beta} \ket{\uparrow_u} - \bar{\alpha} \ket{\downarrow_u}$. Denoting $E$ the probabilistic event ‘Wigner obtains $\uparrow_u$’, do we have $\mathbb{P}_{\mathcal{W}}(E) = \mathbb{P}_{\mathcal{F}}(E)$\ ? Things run as follows: 

\begin{align*}
 \ket{\mathcal{W}_0} \otimes &(c_1 \ket{\uparrow} + c_2 \ket{\downarrow}) \otimes \ket{\mathcal{F}_0} 
\overset{\substack{ \text{Friend measures} \\ \text{along }z}} \longrightarrow \ket{\mathcal{W}_0} \otimes (c_1 \ket{\uparrow \mathcal{F}_{\uparrow}} + c_2 \ket{\downarrow \mathcal{F}_{\downarrow}}) \\
 & \overset{\substack{ \text{Wigner measures} \\ \text{along }u}} \longrightarrow  (\alpha c_1 \ket{\mathcal{W}_{\uparrow_u} \uparrow_u} + \beta c_1 \ket{\mathcal{W}_{\downarrow_u} \downarrow_u}) \otimes \ket{\mathcal{F}_{\uparrow}} + (\bar{\beta} c_2  \ket{\mathcal{W}_{\uparrow_u} \uparrow_u} - \bar{\alpha} c_2 \ket{\mathcal{W}_{\downarrow_u} \downarrow_u}) \otimes \ket{\mathcal{F}_{\downarrow}}. 
\end{align*}
After collapse, the friend's wave function describing Wigner and the spin is either $\alpha \ket{\mathcal{W}_{\uparrow_u} \uparrow_u} + \beta \ket{\mathcal{W}_{\downarrow_u} \downarrow_u}$ or $\bar{\beta}  \ket{\mathcal{W}_{\uparrow_u} \uparrow_u} - \bar{\alpha} \ket{\mathcal{W}_{\downarrow_u} \downarrow_u}$ with probability $\lvert c_1 \rvert^2$ or $\lvert c_2 \rvert^2$ respectively, while Wigner's wave function describing the friend and the spin is (up to a normalization factor) either $\ket{\uparrow_u} \otimes (\alpha c_1 \ket{\mathcal{F}_{\uparrow}} + \bar{\beta} c_2 \ket{\mathcal{F}_{\downarrow}})$ or $\ket{\downarrow_u} \otimes (\beta c_1 \ket{\mathcal{F}_{\uparrow}} - \bar{\alpha} c_2 \ket{\mathcal{F}_{\downarrow}})$ with probability $\lvert \alpha c_1 \rvert^2 + \lvert \bar{\beta} c_2 \rvert^2$ or $\lvert \beta c_1 \rvert^2 + \lvert \bar{\alpha} c_2 \rvert^2$ respectively. Consequently, $\mathbb{P}_{\mathcal{W}}(\uparrow_u) = \lvert \alpha c_1 \rvert^2 + \lvert \bar{\beta} c_2 \rvert^2 = \mathbb{P}_{\mathcal{F}}(\uparrow_u)$. As a general argument, we see that each of the histories split by the friend's measurement is in turn split by Wigner's measurement, so that their respective collapsed wave functions are in fact two marginals extracted from a single classical probability distribution among four possible outcomes displaying no quantum interferences, which are naturally consistent\footnote{Classical probabilities for different observers are consistent because the total probability formula is valid, meaning that (as recalled in \S\ref{basics}) even though the actual value of a variable is not known for an observer, she can still assume that it has a definite value. As an illustration, consider again the trivial example of note \ref{trivial_example}. Suppose that two observers $\mathcal{O}_1$ and $\mathcal{O}_2$ predict the statistics of $x_2$, but only $\mathcal{O}_1$ knows the value of $x_1$. In average for $k \in \llbracket 0 , 3 \rrbracket$, $\mathbb{P}_{\mathcal{O}_1}(x_2=k) = \frac{1}{2} \times 0 +  \frac{1}{2} \times \frac{1}{2} = \frac{1}{4} = \mathbb{P}_{\mathcal{O}_2}(x_2=k)$, but this fundamentally relies on the total probability formula, \textit{i.e.} on the fact that the two possible histories $x_1=0$ and $x_2=1$ did not interfere.}.

\item What happens if the possible histories split by the friend's measurement recombine before Wigner's measurement? For this, suppose now that Wigner can apply any unitary evolution inside the room: 
\[ \frac{1}{\sqrt{2}} (\ket{\uparrow} + \ket{\downarrow})) \otimes \ket{\mathcal{F}_0 } \overset{\substack{ \text{Friend measures} \\ \text{along }z}}{ \underset{U} \longrightarrow} \frac{1}{\sqrt{2}}  (\ket{\uparrow \mathcal{F}_{\uparrow}} + \ket{\downarrow \mathcal{F}_{\downarrow}})  \overset{\text{Wigner applies}}{ \underset{U^{-1}} \longrightarrow}\frac{1}{\sqrt{2}} (\ket{\uparrow} + \ket{\downarrow}) \otimes \ket{\mathcal{F}_0 }. \]
If now Wigner performs a measurement of the spin along $x$, he will get $\uparrow_x$ with certainty, evidence of a superposition of the two possible histories $\uparrow$ and $\downarrow$ along $z$. This means that both histories contribute to the final state\ldots\ even the one that the friend has not experienced! 

Perhaps the most annoying feature of interpretations of QM based on words like ‘worlds’ or ‘histories’ concerns the fate of all the other non-factual histories, which seem very real as long as coherence is preserved, but look so ghostly and unscientific afterwards. But we see here that \textit{when one possible history actually becomes a fact for the friend, all the other non-factual histories are ‘still there’, waiting for re-coherence to interfere again}. If the histories recombine\footnote{Note that for the unitary evolution imposed by Wigner to entail re-coherence, it must decrease the entanglement entropy between $\mathcal{S}$ and $\mathcal{E}$ so, one might say, invert the arrow of time.}, though, for sure the friend doesn't remember what she has observed\footnote{To state this, as well as the second paragraph of \S\ref{definition_measurement}, we have tacitly assumed that the physicist's memories are entirely encoded in her brain's material structure. But do the application of $U^{-1}$ by Wigner really suppresses such immaterial things as memories? If not, it would in theory be possible, for instance, to both observe interferences fringes and know the which-way information in a double slit experiment. However, this hypothesis is not scientifically testable at present because we can't apply precise unitary evolutions on human beings, and will probably never. Personally, I am convinced that spirit is not reducible to matter, but that simple cognitive processes, such that the task of remembering whether the detector was turned on or not, are typically operated by the brain and presumably reducible to it.} (otherwise this would suffice to keep the histories split), so that it is not anymore a fact for her. Wigner and his friend's wave functions describing the spin are now identical, so their predictions are obviously consistent.

Considering that QM is universally valid urges us to accept the idea that \textit{facts have a finite lifetime and that they are relative to an observer's memory} \citep{brukner2020facts}; as \cite{di2021stable} put it: ‘Wigner’s facts are not necessarily his friend’s facts’. It doesn't even make sense to ask whether Wigner would obtain the same outcome as his friend obtained earlier if he measured the spin along $z$ too, because the answer doesn't exist anymore. In \cite{bong2020strong}, the authors constrained this mathematically with a no-go theorem: if QM applies at all scales, then any non-superdeterministic theory satisfying the independence of spacelike separated events can not consistently treat the friend's facts (called ‘observed events’ in the paper) as hidden variables, therefore facts can not exist absolutely. (Note that, for their ‘Local Friendliness (LF) inequalities’ to be violated, they need the superobservers to be able to recombine histories by applying unitary evolutions quite similar to the $U^{-1}$ of our example.) This is reminiscent of Zurek's proposal: ‘I strongly suspect that the ultimate message of quantum theory is that the separation between what exists and what is known to exist – between the epistemic and the ontic – must be abolished’ \citep{zurek2022emergence}. This also reminds the ideas developed in \cite{auffeves2016contexts} and \cite{grangier2018quantum}, in particular the principle according to which ‘In QM, modalities are attributed jointly to the system and to the context’, and finally the ‘background independence’ guiding principle for quantum gravity (see \cite{markopoulou2009new}), in particular the definition given by Dreyer: ‘a theory is background independent if all observations are made by observers \textit{inside} the system’. See also note \ref{note_philo} and the concept of earthly in the work of \cite{latour2015face}. 

\item Suppose now that Wigner is able to perform any measurement of the room's content. Note what a technical feat it must be: for it to be correct, no internal evolution and no external entanglement should occur during the measurement, in particular it must be faster than the already incredibly short (self)decoherence time in position \citep[Table 3.1]{joos1996decoherence}. Thereby, Wigner can make the possible histories of the room interfere without first erasing the friend's facts. Starting again from the state $\ket{\Psi_+} = \frac{1}{\sqrt{2}}  (\ket{\uparrow \mathcal{F}_{\uparrow}} + \ket{\downarrow \mathcal{F}_{\downarrow}})$, let Wigner measure in the $( \ket{\Psi_+} , \ket{\Psi_-})$ basis where $\ket{\Psi_-} = \frac{1}{\sqrt{2}} (\ket{\uparrow \mathcal{F}_{\uparrow}} - \ket{\downarrow \mathcal{F}_{\downarrow}})$, so that he expects to obtain $\ket{\Psi_+}$ with probability 1. On the other hand, the friend, considering herself in the collapsed state $\ket{\uparrow \mathcal{F}_{\uparrow}}$ or $\ket{\downarrow \mathcal{F}_{\downarrow}}$, predicts only a $\frac{1}{2}$ probability to be measured in $\ket{\Psi_+}$, leading to statistical inconsistency. This happens because the friend's collapse was not justified, since Wigner's measurement recombines the two possible histories.

At the end of the measurement, the friend is still in the state $\ket{\Psi_+}$ both for Wigner and herself. How is it like to be superposed? It is probably not a big deal, because (i) everyone is always superposed in most bases, (ii) as seen in 1., Wigner can prepare the friend in a superposed state without even directly interacting with her, (iii) after the measurement, simply by accessing her memories (which may have been rewritten during the measurement, see 4.), the friend splits the histories in the position basis (corresponding for instance to her two possible brain's shapes) and knows which one she is in, therefore she feels nothing more than being in the state $\ket{\uparrow \mathcal{F}_{\uparrow}}$ or $\ket{\downarrow \mathcal{F}_{\downarrow}}$. One is never spatially superposed from one's point of view.

\item Let's finish with the following interesting experiment. Let $(\ket{0}, \ket{1})$ be two possible orthogonal states for the friend in the position basis (for example ‘friend at the right of the room’ and ‘friend at the left of the room’, or ‘friend remembering the digit 0’ and ‘friend remembering the digit 1’), and let Wigner and his friend initially agree together that the latter is in the state $\ket{0}$. Wigner now isolates the friend, measures in the $(\frac{1}{\sqrt{2}}(\ket{0} + \ket{1}), \frac{1}{\sqrt{2}} (\ket{0} - \ket{1}) )$ basis, then opens the door to look at or ask his friend, \textit{i.e.} measures her in the $(\ket{0}, \ket{1})$ basis. Their wave functions describing the friend are initially the same, so here again the statistics are consistent, but they both predict a $\frac{1}{2}$ probability to find the friend in the new state $\ket{1}$. How can this be? This is simply an illustration of item 2.\ (ii) of \S\ref{remarks on QM}, namely that measurements require to interact and perturb the system, all the more as they differ from a position measurement. Here, all the friend's particles have been moved, or the friend's brain has been completely reshaped by the measurement. 
\end{enumerate}

At the end of the day, even for omnipotent observers, QM doesn't lead to inconsistencies at any scale, provided all collapses are justified. It is not really weirder than what it already is for particles, except that it highlights the relativity of facts.

\subsection{What is an observer?} \label{observer}

In \S\ref{definition_measurement}, we used the word ‘observer’ in our definition of the selective measurement, but we haven't specified yet what can be deemed an observer. The problem is raised by \cite{bong2020strong}: ‘the attribution of a “fact” to the friend’s measurement (...) depends on what counts as an “observer” (and what counts as a “measurement”)’, because the experiment they conducted to violate the LF inequalities used photons as friends. So if photons have facts, these are relative, but do we want to consider photons as observers?

This question is not so deep since it is a mere choice of definition; any choice can be acceptable as long as it is consistent with what we have elaborated so far. Does it constrain the definition of an observer? We mainly need observers to have a memory. A few specific degrees of freedom of a particle could suffice, although in this case very few facts can be stored and they are presumably quite short-lived. Note that the verification carried out in \S\ref{consistency} guarantees the consistency even between arbitrary small observers, because the fewer facts can be stored, the fewer collapses are applied and the more alike the different observers' wave functions are, so the fewer inconsistencies can occur. 

That said, what is the point for us to write the wavefunction from a photon's point of view, suppose its facts and apply the collapse for it? Would that even deserve to be called ‘doing physics’? In the end, the most interesting characteristic that we may expect from an observer is perhaps the ability to \textit{communicate} facts and subjective experiences. If we agree on that, it is probably better to keep restricting observers to human beings.

\section*{Conclusion}

The aim of this paper was to investigate the measurement problem of QM in all its ramifications. We argued that the measurement problem is much more a problem of words than a problem of maths. More precisely, it is the challenge of integrating the lessons of the theory of decoherence in the physicist's language, along with accepting the relativity of facts as a fundamental aspect of the universe. We also realized that the usual manichean choice between epistemic and ontic meaning of the wave function is too schematic. To show this, we decomposed the problem into two main questions: the status of the collapse and the emergence of classicality. Examining the first led us to argue that the collapse is not a physical process, but merely a natural consequence of the probabilistic nature of quantum theory. We coined some vocabulary (‘possible histories’, ‘split’, ‘recombine’, ‘fact’\ldots) to speak of the mathematics of QM, in the hope of raising clear pictures in mind with as few counter-intuitive properties as possible. This language was used in the sequel to address the second question: once explained the disappearance of interferences, of quantum correlations and of indeterminism at our scales, we found no reason to believe after all that QM is not universally valid. It also revealed the very special role of the position eigenbasis (both for decoherence and in terms of measurements accessible to humans), shed new lights on the uncertainty principle and highlighted the relativity of facts. The wave function clearly appeared as observer dependent, and we checked that it does not entail any probabilistic inconsistency by revisiting the Wigner's friend thought experiment.

\bibliographystyle{plainnat}
\bibliography{Biblio_measurement}

\begin{thebibliography}{28}
\providecommand{\natexlab}[1]{#1}
\providecommand{\url}[1]{\texttt{#1}}
\expandafter\ifx\csname urlstyle\endcsname\relax
  \providecommand{\doi}[1]{doi: #1}\else
  \providecommand{\doi}{doi: \begingroup \urlstyle{rm}\Url}\fi

\bibitem[Auff{\`e}ves and Grangier(2016)]{auffeves2016contexts}
Alexia Auff{\`e}ves and Philippe Grangier.
\newblock Contexts, systems and modalities: a new ontology for quantum
  mechanics.
\newblock \emph{Foundations of Physics}, 46\penalty0 (2):\penalty0 121--137,
  2016.
\newblock \doi{10.1007/s10701-015-9952-z}.

\bibitem[Auff{\`e}ves and Grangier(2018)]{grangier2018quantum}
Alexia Auff{\`e}ves and Philippe Grangier.
\newblock What is quantum in quantum randomness?
\newblock \emph{Philosophical Transactions of the Royal Society A:
  Mathematical, Physical and Engineering Sciences}, 376\penalty0
  (2123):\penalty0 20170322, 2018.
\newblock \doi{10.1098/rsta.2017.0322}.

\bibitem[Bassi et~al.(2013)Bassi, Lochan, Satin, Singh, and
  Ulbricht]{bassi2013models}
Angelo Bassi, Kinjalk Lochan, Seema Satin, Tejinder~P Singh, and Hendrik
  Ulbricht.
\newblock Models of wave-function collapse, underlying theories, and
  experimental tests.
\newblock \emph{Reviews of Modern Physics}, 85\penalty0 (2):\penalty0 471,
  2013.
\newblock \doi{10.1103/RevModPhys.85.471}.

\bibitem[Bedingham et~al.(2014)Bedingham, D{\"u}rr, Ghirardi, Goldstein,
  Tumulka, and Zangh{\`\i}]{bedingham2014matter}
Daniel Bedingham, Detlef D{\"u}rr, GianCarlo Ghirardi, Sheldon Goldstein,
  Roderich Tumulka, and Nino Zangh{\`\i}.
\newblock Matter density and relativistic models of wave function collapse.
\newblock \emph{Journal of Statistical Physics}, 154:\penalty0 623--631, 2014.
\newblock \doi{10.1007/s10955-013-0814-9}.

\bibitem[Bell(1982)]{bell1982impossible}
John~S Bell.
\newblock On the impossible pilot wave.
\newblock \emph{Foundations of Physics}, 12\penalty0 (10):\penalty0 989--999,
  1982.
\newblock \doi{10.1007/BF01889272}.

\bibitem[Bong et~al.(2020)Bong, Utreras-Alarc{\'o}n, Ghafari, Liang, Tischler,
  Cavalcanti, Pryde, and Wiseman]{bong2020strong}
Kok-Wei Bong, An{\'\i}bal Utreras-Alarc{\'o}n, Farzad Ghafari, Yeong-Cherng
  Liang, Nora Tischler, Eric~G Cavalcanti, Geoff~J Pryde, and Howard~M Wiseman.
\newblock A strong no-go theorem on the {W}igner’s friend paradox.
\newblock \emph{Nature Physics}, 16\penalty0 (12):\penalty0 1199--1205, 2020.
\newblock \doi{10.1038/s41567-020-0990-x}.

\bibitem[Brukner(2020)]{brukner2020facts}
{\v{C}}aslav Brukner.
\newblock Facts are relative.
\newblock \emph{Nature Physics}, 16\penalty0 (12):\penalty0 1172--1174, 2020.
\newblock \doi{10.1038/s41567-020-0984-8}.

\bibitem[Carlesso et~al.(2022)Carlesso, Donadi, Ferialdi, Paternostro,
  Ulbricht, and Bassi]{carlesso2022present}
Matteo Carlesso, Sandro Donadi, Luca Ferialdi, Mauro Paternostro, Hendrik
  Ulbricht, and Angelo Bassi.
\newblock Present status and future challenges of non-interferometric tests of
  collapse models.
\newblock \emph{Nature Physics}, 18\penalty0 (3):\penalty0 243--250, 2022.

\bibitem[Chapman et~al.(1995)Chapman, Hammond, Lenef, Schmiedmayer, Rubenstein,
  Smith, and Pritchard]{chapman1995photon}
Michael~S Chapman, Troy~D Hammond, Alan Lenef, J{\"o}rg Schmiedmayer, Richard~A
  Rubenstein, Edward Smith, and David~E Pritchard.
\newblock Photon scattering from atoms in an atom interferometer: coherence
  lost and regained.
\newblock \emph{Physical Review Letters}, 75\penalty0 (21):\penalty0 3783,
  1995.
\newblock \doi{10.1103/PhysRevLett.75.3783}.

\bibitem[Dewdney et~al.(1993)Dewdney, Hardy, and Squires]{dewdney1993late}
Chris Dewdney, Lucien Hardy, and Euan~J Squires.
\newblock How late measurements of quantum trajectories can fool a detector.
\newblock \emph{Physics Letters A}, 184\penalty0 (1):\penalty0 6--11, 1993.
\newblock \doi{10.1016/0375-9601(93)90337-Y}.

\bibitem[Di~Biagio and Rovelli(2021)]{di2021stable}
Andrea Di~Biagio and Carlo Rovelli.
\newblock Stable facts, relative facts.
\newblock \emph{Foundations of Physics}, 51\penalty0 (1):\penalty0 1--13, 2021.
\newblock \doi{10.1007/s10701-021-00429-w}.

\bibitem[Dove(1996)]{dove1996explicit}
C.~Dove.
\newblock Explicit wavefunction collapse and quantum measurement.
\newblock \emph{Ph.D. thesis, Department of Mathematical Sciences, University
  of Durham}, 1996.

\bibitem[Englert et~al.(1992)Englert, Scully, S{\"u}ssmann, and
  Walther]{englert1992surrealistic}
Berthold-Georg Englert, Marian~O Scully, Georg S{\"u}ssmann, and Herbert
  Walther.
\newblock Surrealistic bohm trajectories.
\newblock \emph{Zeitschrift f{\"u}r Naturforschung A}, 47\penalty0
  (12):\penalty0 1175--1186, 1992.
\newblock \doi{10.1515/zna-1992-1201}.

\bibitem[Heisenberg(1973)]{heisenberg1973development}
Werner Heisenberg.
\newblock Development of concepts in the history of quantum theory.
\newblock In \emph{The physicist’s conception of nature}, pages 264--275.
  Springer, 1973.
\newblock \doi{10.1007/978-94-010-2602-4}.

\bibitem[Joos(1996)]{joos1996decoherence}
Erich Joos.
\newblock Decoherence through interaction with the environment.
\newblock In \emph{Decoherence and the appearance of a classical world in
  quantum theory}, pages 35--136. Springer, 1996.
\newblock \doi{10.1007/978-3-662-03263-3}.

\bibitem[Latour(2015)]{latour2015face}
Bruno Latour.
\newblock \emph{Face {\`a} Ga{\"\i}a: huit conf{\'e}rences sur le nouveau
  r{\'e}gime climatique}.
\newblock Emp{\^e}cheurs de penser rond, 2015.
\newblock \doi{10.4000/lectures.19763}.

\bibitem[Markopoulou(2009)]{markopoulou2009new}
Fotini Markopoulou.
\newblock New directions in background independent quantum gravity.
\newblock \emph{Approaches to quantum gravity}, pages 129--149, 2009.
\newblock \doi{10.1017/CBO9780511575549}.

\bibitem[Peterson(15 april 2022)]{interviewpenrose}
Jordan~B Peterson.
\newblock Why quantum mechanics is an inconsistent theory | {R}oger {P}enrose
  \& {J}ordan {P}eterson, 15 april 2022.
\newblock {\url{https://www.youtube.com/watch?v=TSBOBJsdEuY}}.

\bibitem[Schlosshauer(2005)]{schlosshauer2005decoherence}
Maximilian Schlosshauer.
\newblock Decoherence, the measurement problem, and interpretations of quantum
  mechanics.
\newblock \emph{Reviews of Modern physics}, 76\penalty0 (4):\penalty0 1267,
  2005.
\newblock \doi{10.1103/RevModPhys.76.1267}.

\bibitem[Schlosshauer et~al.(2013)Schlosshauer, Kofler, and
  Zeilinger]{schlosshauer2013snapshot}
Maximilian Schlosshauer, Johannes Kofler, and Anton Zeilinger.
\newblock A snapshot of foundational attitudes toward quantum mechanics.
\newblock \emph{Studies in History and Philosophy of Science Part B: Studies in
  History and Philosophy of Modern Physics}, 44\penalty0 (3):\penalty0
  222--230, 2013.
\newblock \doi{10.1016/j.shpsb.2013.04.004}.

\bibitem[Sivasundaram and Nielsen(2016)]{sivasundaram2016surveying}
Sujeevan Sivasundaram and Kristian~Hvidtfelt Nielsen.
\newblock Surveying the attitudes of physicists concerning foundational issues
  of quantum mechanics.
\newblock \emph{arXiv preprint arXiv:1612.00676}, 2016.
\newblock \doi{10.48550/arXiv.1612.00676}.

\bibitem[Soulas()]{soulas2023logical}
Antoine Soulas.
\newblock Logical implications between fundamental properties of relativistic
  quantum theories.
\newblock \emph{arXiv preprint}.

\bibitem[Soulas(2023)]{soulas2023decoherence}
Antoine Soulas.
\newblock Decoherence as a high-dimensional geometrical phenomenon.
\newblock \emph{arXiv preprint}, 2023.
\newblock doi : 10.48550/arXiv.2302.04148.

\bibitem[Tumulka(2006)]{tumulka2006relativistic}
Roderich Tumulka.
\newblock A relativistic version of the ghirardi--rimini--weber model.
\newblock \emph{Journal of Statistical Physics}, 125:\penalty0 821--840, 2006.
\newblock \doi{10.1007/s10955-006-9227-3}.

\bibitem[Wittgenstein(1953)]{wittgenstein1953philosophical}
Ludwig Wittgenstein.
\newblock \emph{Philosophical investigations}.
\newblock Basil Blackwell, Oxford, 1953.
\newblock ISBN 9780631119005.

\bibitem[Zurek(1982)]{zurek1982environment}
Wojciech~Hubert Zurek.
\newblock Environment-induced superselection rules.
\newblock \emph{Physical review D}, 26\penalty0 (8):\penalty0 1862, 1982.
\newblock \doi{10.1103/PhysRevD.26.1862}.

\bibitem[Zurek(2003)]{zurek2003decoherence}
Wojciech~Hubert Zurek.
\newblock Decoherence, einselection, and the quantum origins of the classical.
\newblock \emph{Reviews of modern physics}, 75\penalty0 (3):\penalty0 715,
  2003.
\newblock \doi{10.1103/RevModPhys.75.715}.

\bibitem[Zurek(2022)]{zurek2022emergence}
Wojciech~Hubert Zurek.
\newblock Emergence of the classical world from within our quantum universe.
\newblock In \emph{From Quantum to Classical}, pages 23--44. Springer, 2022.
\newblock \doi{10.1007/978-3-030-88781-0_2}.

\end{thebibliography}
\end{document}